\input{aipcheck}

\documentclass[
    ,final            % use final for the camera ready runs
  ]
  {aipproc}

\layoutstyle{8x11single}

\begin{document}

\title{Lessons from LHC elastic and diffractive data}

\classification{
                \texttt{13.85.Dz, 13.85.Lg, 11.80.Gw}}

\author{A.D. Martin, V.A. Khoze and M.G. Ryskin}{
  address={Institute for Particle Physics Phenomenology, Durham University, Durham, DH1 3LE}
}

\begin{abstract}
 In the light of LHC data, we discuss the global description of all high energy elastic and diffractive data, using a one-pomeron model, but including multi-pomeron interactions.  The LHC data indicate the need of a $k_t(s)$ behaviour, where $k_t$ is the gluon transverse momentum along the partonic ladder structure which describes the pomeron. We also discuss tensions in the data, as well as  the $t$ dependence of the slope of $d\sigma_{\rm el}/dt$ in the small $t$ domain.
\end{abstract}

\maketitle

\section{Description of `soft' data before LHC}

 The KMR \cite{KMR-s3} approach to describe high-energy elastic and diffractive $pp$ data uses the framework of Regge pomeron theory, which includes non-enhanced eikonal-like multi-pomeron interactions together with the Good-Walker formalism \cite{GW} for diffractive eigenstates to describe elastic scattering and low-mass proton dissociation.
 High-mass dissociation is described by including multi-pomeron vertices which account for the rescattering of intermediate partons. The procedure is sketched below:
 
\begin{figure} [h]
\includegraphics[clip=true,trim=1.1cm 0.1cm 0cm 6cm,height=4.5cm]{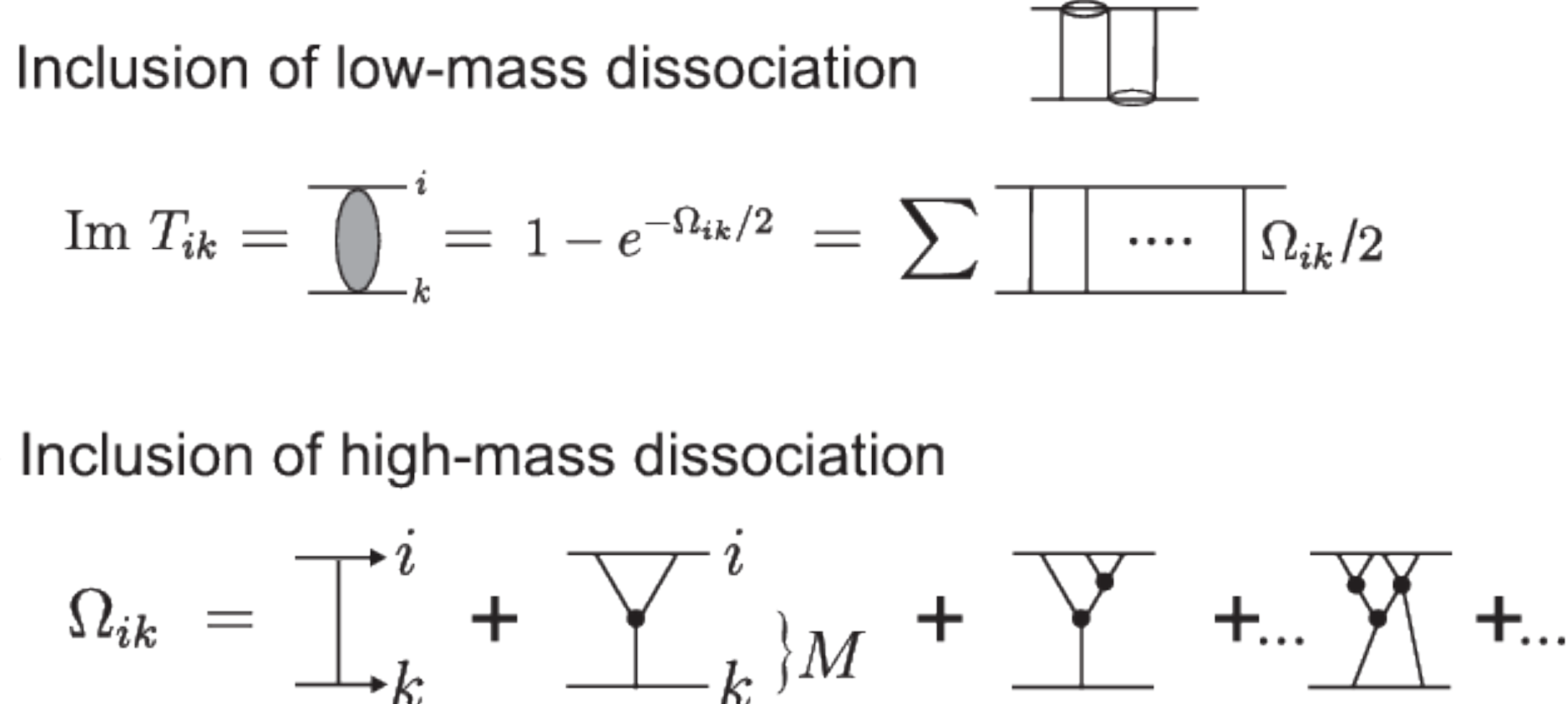}
\label{fig:AA}
\caption{The first equation sketches the multi-channel eikonal expression of the amplitude for the scattering of diffractive eigenstates $i,k$ -- states that are the linear combinations of $|p\rangle,|p^*\rangle,...$ which undergo elastic-type scattering. In this way we can describe elastic scattering and low-mass dissociation. Next we show the multi-pomeron diagrams, that involve the coupling $g^m_n$ of $m$ to $n$ pomerons, which allow for high-mass dissociation -- the simplest being the triple-pomeron diagram describing proton dissociation into the high-mass $M$ system. All the `$t$-channel lines' represent pomeron exchange, which in pQCD have a ladder-type structure -- we speak of the `hard' or `BFKL' pomeron.}
\end{figure}

 Contrary to conventional Regge theory, where it was assumed that all transverse momenta, $k_t$, are limited, we try to match the perturbative QCD and Regge Field Theory (RFT) approaches.  We start with the BFKL hard pomeron, where the parton $k_t$ may increase or decrease at each step of the log$(1/x)$ evolution (along the `ladder'). However, stronger absorption of the low $k_t$ partons, described by multi-pomeron vertices, leads to a growth of $\langle  k_t \rangle$ with energy. This justifies the pQCD approach, and. moreover, is consistent with the increase of the infrared cutoff with energy, $k_{t~{\rm min}} \propto s^{0.12}$, which is needed to describe the spectra of secondaries by the Monte Carlo generators \cite{MC1,MC2} 

The growth of  $k_{t~{\rm min}}$ with energy is generated by the evolution of the hard pomeron. It is the appearance of this dynamical cutoff which allows us to extrapolate the predictions of pQCD from the large $k_t$ region to the soft domain. In more detail, at LO, the parton density is described by the BFKL equation
\begin{equation}
\frac{\partial f(y,k_t)}{\partial y}=\alpha_s\int d^2k'_t~ K(k_t,k'_t)~f(y,k'_t)
\label{eq:bfkl}
\end{equation}
with $y=$log$(1/x)$. The evolution is distorted by the probability  of absorption of an intermediate gluon, which in RFT, is described by multi-pomeron vertices. To include this, we multiply the BFKL kernel $K$ by the factor\footnote{The absorptive factor exp$(-\lambda \Omega/2)$ includes not only the three-pomeron, but also the higher multi-pomeron vertices, coming from the expansion of the exponent in powers of $\lambda \Omega$. However, to be consistent with the AGK cutting rules \cite{AGK} we replace exp$(-\lambda \Omega/2)$ by [1$-$exp$(-\lambda \Omega)]/\lambda \Omega$, which produces a common eikonal-like form, $g^m_n \propto (\lambda \Omega)^{n+m}$ coupling $m$ to $n$ pomerons.} exp$(-\lambda \Omega)$ where $\Omega$ is the optical density of the incoming proton.  Since the optical density is proportional to the density of partons, we can write the evolution equations as
\begin{equation}
\frac{\partial \Omega_k(y)}{\partial y}~=~\alpha_s\int  d^2k'_t  ~{\rm exp}(-\lambda[\Omega_k(y)+\Omega_i(y')]/2)~K(k_t,k'_t)~\Omega_k(y).
\label{e15}
\end{equation}
\begin{equation}
\frac{\partial \Omega_i(y')}{\partial y'}~=~\alpha_s\int d^2k'_t  ~{\rm exp}(-\lambda[\Omega_i(y')+\Omega_k(y)]/2)~K(k_t,k'_t)~\Omega_i(y').
\label{e16}
\end{equation}
The factor $\lambda$ accounts for the difference of the proton opacity `measured' by the other proton in comparison with that `measured' by the intermediate gluon in the BFKL evolution. The crucial point is the dependence of $\lambda$ on the $k_t$ of the current gluon
\begin{equation}
\lambda \propto \alpha_s(k_t)\Theta(k'_t-k_t)/k^2_t,
\label{eq:lam}\end{equation}
where the $\Theta$ function arises since a large size pomeron component  does not interact with a small size pomeron, which for $k'_t \ll k_t$ looks like a colourless point-like object. The last factor in (\ref{eq:lam}) reflects the dependence on $\sigma^{\rm abs} \sim 1/k_t^2$.

Note that we account for absorption by both the beam and target proton diffractive eigenstates $i$ and $k$. So instead of the single evolution equation (\ref{eq:bfkl}), we now have a pair of equations. One evolving $\Omega_k$ up in $y$ from the target proton eigenstate $k$, the other evolving $\Omega_i$ down from the beam $i$. where $y'=$log$(s/k_t^2)-y$. The equations can be solved by iteration. The resulting $k_t$ behaviour of the high-energy amplitude can now be safely continued down into the soft domain.

In \cite{KMR-s3} we started with a BFKL pomeron which, at LO, has a large intercept, $\alpha_{\rm pom}(0)\simeq 1.5$, However, when we include the absorptive corrections and the kinematic constraint the effective intercept drops to about 1.12; and we were able to describe all the available pre-LHC, high-energy elastic and diffractive data; namely, the energy behaviour of $\sigma_{\rm tot}$, $d\sigma_{\rm el}/dt,~\sigma^{{\rm low}M}_{\rm SD}$, and the probability to observe Large Rapidity Gap (LRG) events, that is high $M$ dissociation.

\section{Surprises from the LHC}
What are the predictions for the LHC?~  For 7 TeV, we predicted that $\sigma_{\rm tot}=88$ mb, the elastic slope, $B_{\rm el}(0)=18.5 ~{\rm GeV}^{-2}$ and $\sigma^{{\rm low}M}_{\rm SD}\simeq 6$ mb.  These are to be compared with the TOTEM measurements of $\sigma_{\rm tot}=98.6 \pm 2.2$ mb, $B_{\rm el}(0)=19.9\pm 0.3 ~{\rm GeV}^{-2}$ \cite{TOTEMtot13a} and $\sigma^{{\rm low}M}_{\rm SD}= 2.6 \pm 2.2$ mb \cite{TOTEMlowmass}. The corresponding ATLAS(ALFA) \cite{ALFA} measurements are $\sigma_{\rm tot}=95.35\pm 1.3$ mb and $B_{\rm el}(0)=19.73\pm0.24~{\rm GeV}^{-2}$.  Our predictions, along with the other attempts \cite{GLM,Ost11,KaidPog} at global descriptions of pre-LHC data, fail to describe the LHC data. 

We may summarize the deficiencies as follows. On the one hand, the total cross section and the elastic slope appear to increase faster than expected from the description of the data at lower collider energies. On the other hand, the probability of low-mass dissociation, and of the diffractive cross section $d\sigma/d\eta$ of high-mass dissociation in the central rapidity interval are smaller at the LHC than expected.

Let us highlight one deficiency.  Low-mass diffractive dissociation has been estimated in three experiments; CERN-ISR\footnote{The relevant experimental references are listed in \cite{KMRC73}.} in the range 31$-$62.5 GeV, ~UA4 at 546 GeV \cite{UA4lowmass}, ~TOTEM at 7 TeV \cite{TOTEMlowmass}, giving, respectively, (in mb)
\begin{equation}
\sigma^{{\rm low}M}_{\rm SD}~/~\sigma_{\rm el}~~~~=~~~~2-3/7,~~~~~~3/12,~~~~~~2.6/25.
\end{equation}
Naively, since both processes are driven by pomeron exchange, we would have expected the ratios to be equal; that is more or less independent of the collider energy. Yet the measurements at the LHC are far smaller than expected based on conventional Regge theory.

There is something missing in the model.

\section{The missing ingredient}

Now, the triple- and multi-pomeron couplings, and the coupling of the pomeron to the GW eigenstates of the proton have no reason to be independent of energy, as is assumed in the models. In fact we should expect them to depend on energy and on the position of the multi-pomeron vertex in rapidity. Indeed these couplings are dimensionful quantities, and the parameter which controls their values is the $k_t$ of the corresponding parton.  This $k_t$ was limited in old RFT and so the vertices were assumed to be constant. Now, thanks to BFKL diffusion in log$k_t$ space, and the stronger absorption of low $k_t$ partons, we see the typical value of $k_t$ increases with energy and with the increase of the available rapidity interval of the evolution. 

Therefore it is natural to expect a decrease of the pomeron couplings as energy increases.  For high-mass dissociation it results in a lower diffractive cross section, $d\sigma/d\eta$, in the central rapidity interval. This effect was predicted in \cite{KMR-s3} and observed in the preliminary TOTEM data \cite{TOTEMhm,TOTEMdiff14}.

Let us return to the couplings, $\gamma_i$ of the pomeron to the diffractive eigenstates, $\phi_i$ of the proton.  They are driven by the transverse separation $\langle r_i^{\rm parton}\rangle$ of the partons
in the $\phi_i$ states. However, the $\gamma_i$'s are also influenced by the transverse size of the pomeron ($\propto 1/k_{\rm pom}$) when it becomes smaller than $\langle r_i^{\rm parton}\rangle \propto 1/k_i$.  Thus it is natural to take
\begin{equation}
\gamma_i \propto 1/(k_{\rm pom}^2+k_i^2)~~~~~~~~~{\rm where}~k_{\rm pom}^2~{\rm is~parametrised~by}~~~~~~~~~k_{\rm pom}^2=k_0^2s^{0.28}.
\end{equation}
It follows, that as $s \to \infty$, all the couplings $\gamma_i$ become equal, $\gamma_i \propto 1/k^2_{\rm pom}$, and so the dispersion, induced by the interaction with the pomeron, decreases, leading to a much smaller probability of low mass dissociation. Simultaneously, due to the smaller dispersion, the whole absorptive effect becomes weaker, allowing a faster growth of the total cross section within the energy interval in which the dispersion disappears.

Including this effect (that is, energy dependent coupling $\gamma_i$) we obtain \cite{kmr2} a good description of all high-energy elastic and diffractive data, including the LHC measurements. We see, for example, from Table 1, that the deficiencies in the description of TOTEM data, that were mentioned before, are removed. 
\begin{table} [h]
\begin{tabular}{lccccc} 
\hline
 & & \tablehead{1}{r}{b}{Tevatron}
  & \tablehead{1}{r}{b}{LHC (7 TeV)}
  & \tablehead{1}{r}{b}{100 TeV}
  & \tablehead{1}{r}{b} {TOTEM (7 TeV)}   \\
\hline
$\sigma(\rm tot)$ & mb & 77 & 98.7    & 166 & $  98.6 \pm 2.2  $ \\
$B_{\rm el}(0)$ & GeV$^{-2}$ & 16.8 & 19.7 & 29.4 & $19.9\pm 0.3$\\
$\sigma_{\rm SD}$(low $M$) & mb & 3.4 & 3.6 & 2.7 & $2.6\pm 2.2$\\
\hline
\end{tabular}
\caption{The energy behaviour of observables with the $k_t(s)$ effect included \cite{kmr2}. We see that the effect brings the model into agreement with the TOTEM data \cite{TOTEMtot13a,TOTEMlowmass}.}
\label{tab:a}
\end{table}

\section{Tension in high-mass dissociation data}
\begin{figure} [h]
\includegraphics[clip=true,trim=1cm 0.2cm 1cm 10cm,height=7cm]{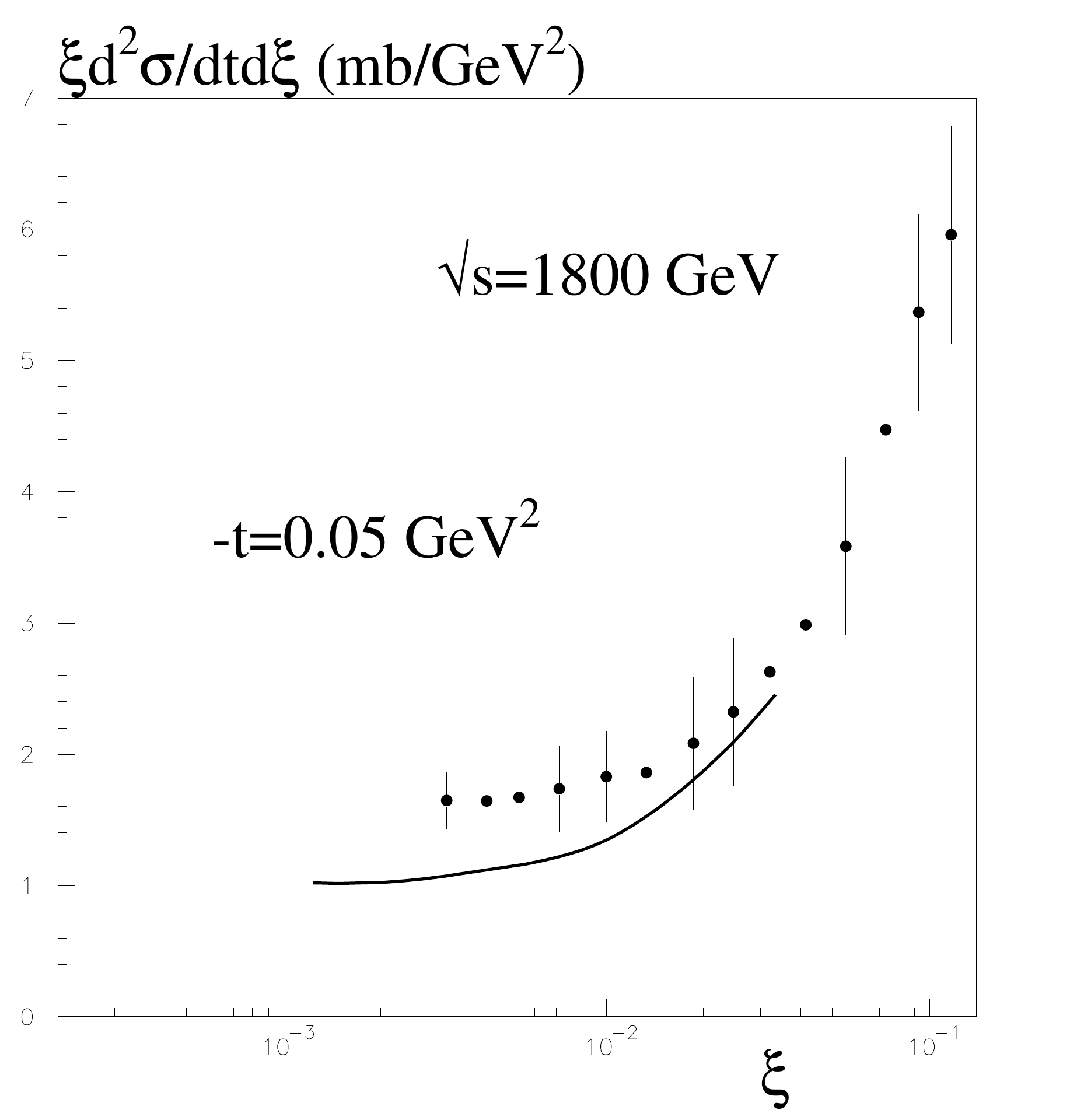}\rule{0cm}{0cm}
\includegraphics[clip=true,trim=1cm 0.2cm 1cm 10cm,height=7cm]{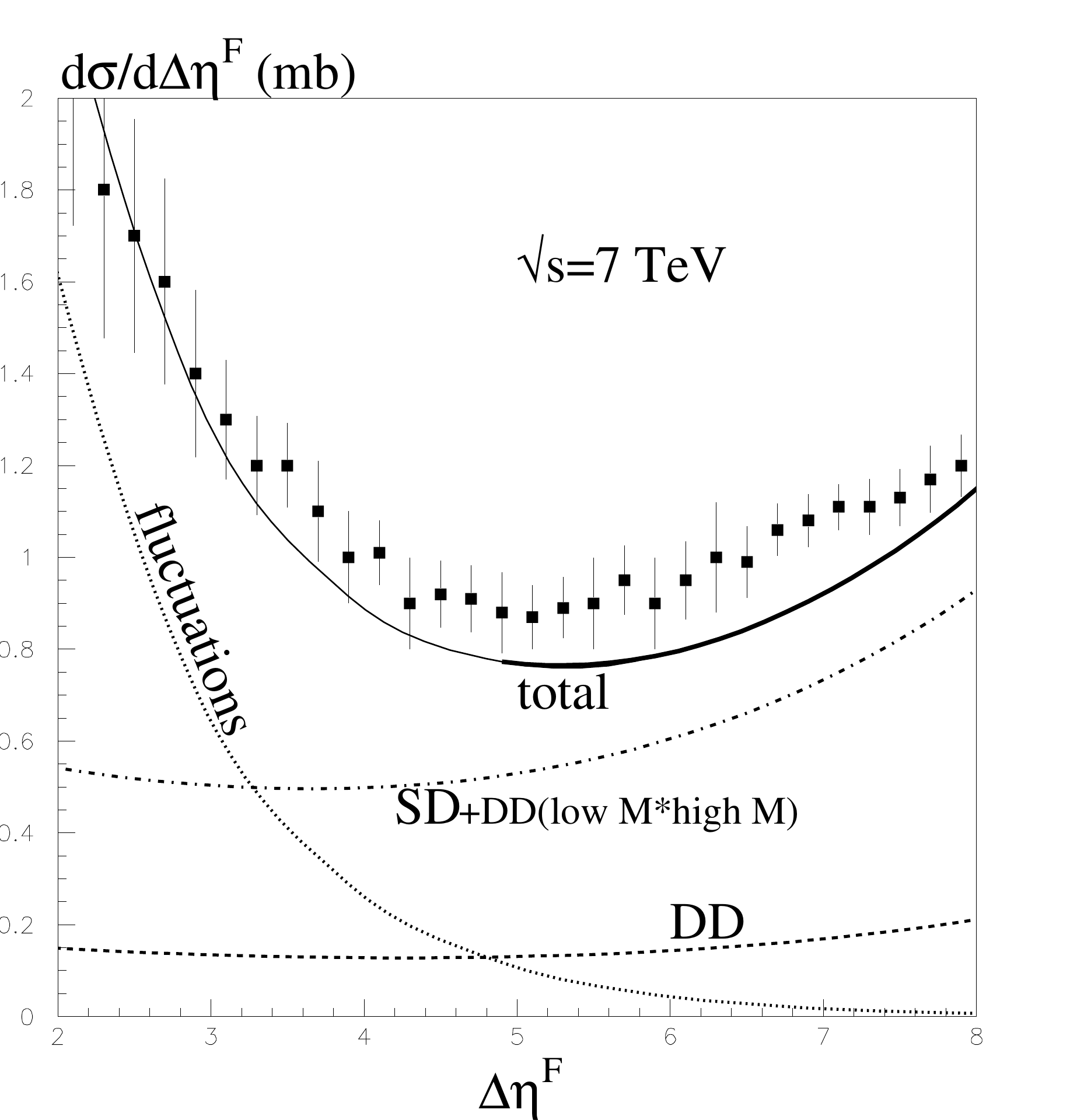}
\label{fig:2}
\caption{(a) The comparison of the model \cite{kmr2} with data for single proton dissociation measured by the CDF collaboration, given in \cite{GM} but not including a normalisation uncertainty of about 10-15\%. The inclusion of the secondary Reggeon contribution RRP is responsible for the rise of the curve as $\xi$ increases. (b) The ATLAS \cite{atl} measurements of the inelastic cross section differential in rapidity gap size $\Delta\eta^F$ for particles with $p_T>200$ MeV. Events with small gap size ($\Delta\eta^F < 5$) may have a non-diffractive component which arises from fluctuations in the hadronization process \cite{FZ}. This contribution increases as $\Delta\eta^F$ decreases (or if a larger $p_T$ cut is used \cite{FZ,atl}).  
The data with $\Delta\eta^F >5$ are dominantly of diffractive origin, and are compared with the `global' diffractive model \cite{kmr2}. The DD contribution of events where both protons dissociate, but the secondaries produced by one proton go into the beam pipe and are not seen in the calorimeter, is shown by the dashed curve. The figures are taken from \cite{kmr2}.}
\end{figure}

Although TOTEM have made the most detailed observations of high-mass single proton dissociation in high energy $pp$ collisions, the  `global' two-channel eikonal model of \cite{kmr2} was tuned to simultaneously describe the TOTEM data {\it together} with earlier measurements of single dissociation. 
Formally the diffractive dissociation data from different groups do not contradict each other, since they are measured under different conditions. However, global fits of all diffractive data reveal some tension between the data sets \cite{kmr2,Ost14}. 
 Indeed, it is not easy to reconcile the preliminary TOTEM measurements \cite{TOTEMhm,TOTEMdiff14} for $\sigma_{\rm SD}$ with (a) the `CDF' measurement of single dissociation and (b) the yield of Large Rapidity Gap events observed by  ATLAS and CMS.  Basically the global description of \cite{kmr2} overestimates the TOTEM data and underestimates the CDF \cite{GM}, ATLAS \cite{atl} and CMS \cite{CMSdiff} data. See Fig. \ref{fig:2} for the CDF and ATLAS comparison with the model.
For high-mass dissociation the preliminary TOTEM measurement, integrated over a mass interval (8$-$350) GeV, is $\sigma_{\rm SD}=3.3$ mb, while CMS find $\sigma_{\rm SD}=4.3$ mb over the comparable (12$-$394) GeV mass interval; and the model gives the intermediate value of 4.0 mb.
\begin{figure} [h]
\includegraphics[clip=true,trim=1cm 0.2cm 1cm 10cm,height=7cm]{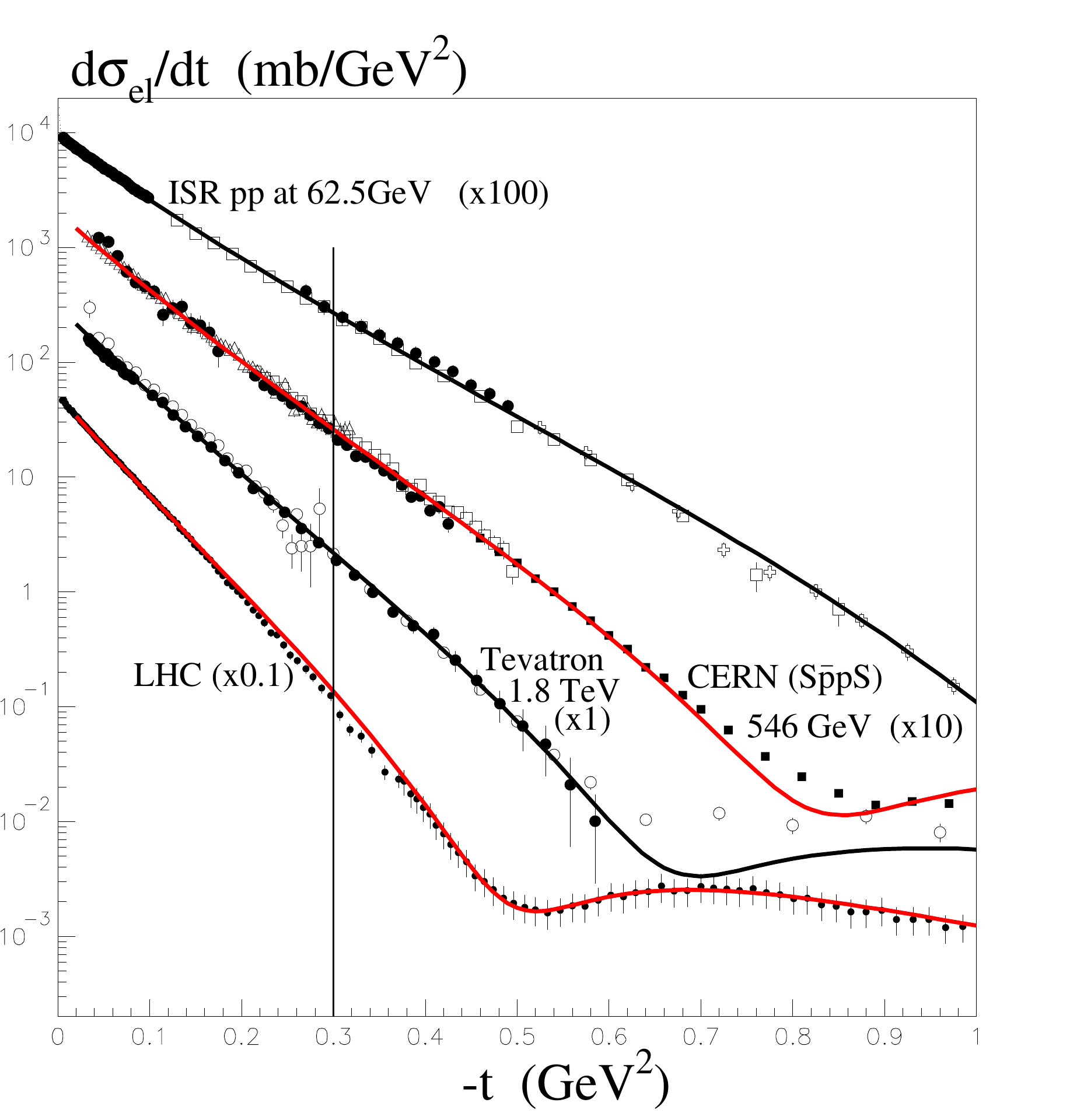}\rule{0cm}{0cm}
\includegraphics[clip=true,trim=1cm 0.2cm 1cm 10cm,height=7cm]{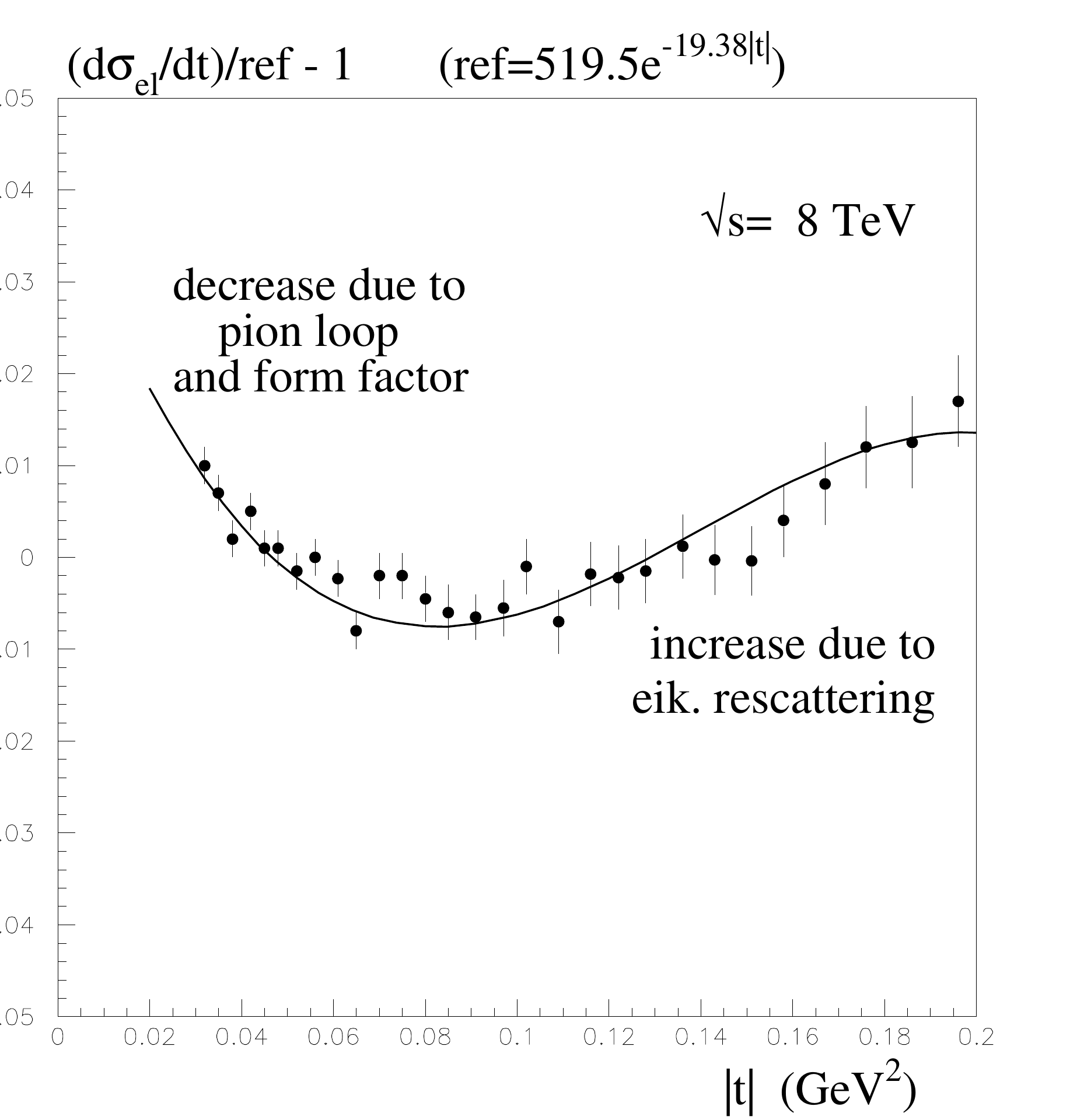}
\label{fig:3}
\caption{(a) The references to the elastic data can be found, for example, in \cite{kmrBslope}; the LHC points are at 7 TeV; the curves are from \cite{kmr2}. (b) The deviation of $d\sigma_{\rm el}/dt$ at 8 TeV from a pure exponential; the data points are taken from \cite{TOTEMt,TOTEMdiff14} and the curve is the prediction from \cite{kmr2}.}
\end{figure}

\section{Elastic scattering and the $B$ slope versus $t$}

The left diagram in Fig.\ref{fig:3} shows the description obtained in \cite{kmr2} of the data for the elastic differential cross section at various collider energies. 
In the global fit we find a two-channel eikonal is sufficient to describe the low-mass dissociation data. However, to describe the elastic data in the region of the diffractive dip it is crucial to include the real part of the amplitude -- which was calculated using dispersion relations. The detailed description of the data in Fig.\ref{fig:3}(a) relies on the parameters specifying the $t$ dependence of the pomeron couplings to the diffractive eigenstates.  These form factors turn out to have approximately Orear-like forms \cite{orear}, $F_i(t) \sim ~$exp$(-b_i\sqrt{|t|})$.

Let us concentrate on the small $t$ domain, $|t|<0.2 ~{\rm GeV}^2$. At first sight, it appears that the small $|t|$ data are well described by an exponent form
\begin{equation}
d\sigma_{\rm el}/dt=d\sigma_{\rm el}/dt\big|_{t=0}\cdot\exp(Bt)
\end{equation}
where the slope parameter is a constant independent ot $t$, but increases with collider energy. However a more detailed look shows that $B$ has a complicated behaviour. Indeed there is no reason why $B$ should be constant. This is an approximation. The nearest singularity at $t=4m^2_\pi\simeq 0.08$ GeV$^2$ corresponds to the production (in $t$-channel) of a pair of pions, and this threshold leads to a non-linear dependence of the pomeron trajectory, $\alpha_{\rm pom}(t)$, on $t$.  Besides this, we have the $t$ dependence of the pomeron couplings which are described by the Orear-like form factors.  Both effects lead to a decrease of the slope $B$ at relatively low $|t|$. On the other hand the absorptive effects cause an increase of the slope especially when $|t|$ starts to approach the position of the diffractive dip. Such behaviour is indeed observed. The diagram on the right of Fig.\ref{fig:3} shows the deviation of the elastic cross section from a pure exponential fit (denoted by `ref' on the top of the diagram) observed in the (preliminary) TOTEM data at 8 TeV \cite{TOTEMt,TOTEMdiff14}.  We see the prediction of \cite{kmr2} is consistent with the data.  Finally, in \cite{kmrBslope}, we note that the deviation from a pure exponential form is expected to be more pronounced as the collider energy increases from the CERN S$p\bar{p}$S, to the Tevatron, to the LHC and beyond.

\section{Conclusion}
Two of the main surprises of the LHC elastic and diffractive data are that the total $pp$ cross section is larger than expected and that low-mass dissociation is smaller than predicted. We argue that there is a physics reason for this. In the energy regime embracing the Tevatron and LHC colliders, the coupling $\gamma_i$ of the pomeron to the diffractive eigenstate $\phi_i$ of proton go, from being driven mainly by the mean transverse separation of the partons in the $\phi_i$ state, to being mainly controlled by the transverse size of the pomeron -- which decreases as the collider energy increases. The $\gamma_i$, therefore, tend to a common value as the energy increases. Thus the dispersion, and hence the probability of proton low-mass dissociation, decreases. Moreover the two-channel eikonal becomes closer to a one-channel eikonal, and the absorption becomes smaller, which speeds up the growth of the total cross section in this particular energy interval. We called this behaviour of the $\gamma_i$, the  $k_t(s)$ effect.

Although the $k_t$ of the parton is not directly observable, one way to check this effect is to measure the $p_t$ distributions of mesons containing a heavy $c$ or $b$ quark, or better to measure $D\bar{D}$ or $B\bar{B}$ meson pairs, as a function of energy \cite{kty}.

\bibliographystyle{aipproc}   % if natbib is available

\bibliography{F}

\end{document}